\documentclass[aps, prd, amsmath, floats,floatfix, twocolumn, superscriptaddress,
nofootinbib, showpacs]{revtex4-1}

\usepackage{amssymb}
\usepackage{amsmath}
\usepackage{verbatim}
\usepackage{mathrsfs}
\usepackage{amsfonts}
\usepackage{latexsym}
\usepackage{epsfig}
\usepackage{color}
\usepackage{graphicx,subfigure}
\usepackage{units}
\usepackage{slashbox}

\begin{document}

\title{Note on scalars, perfect fluids, constrained field theories, and all that}

\author{Alberto Diez-Tejedor}
\affiliation{Departamento de F\'isica, Divisi\'on de Ciencias e Ingenier\'ias, Campus Le\'on, 
Universidad de Guanajuato, Le\'on 37150, M\'exico}

\date{\today}

\begin{abstract} 
The relation of a scalar field with a perfect fluid has generated some debate along the last few years. In this paper we argue that shift-invariant scalar fields can describe accurately the potential flow of an isentropic perfect fluid, but, in general, the identification is possible only for a finite period of time. After that period in the evolution the dynamics of the scalar field and the perfect fluid branch off. The Lagrangian density for the velocity-potential can be read directly from the expression relating the pressure with the Taub charge and the entropy per particle in the fluid, whereas the other quantities of interest can be obtained from the thermodynamic 
relations. 
\end{abstract}


\maketitle

\section{Introduction}

Recently many papers have addressed the question: can we identify a scalar field with the potential flow of a perfect fluid? 
For a representative sample of these works please see Refs.~\cite{DiezTejedor:2005fz}. Here we will concentrate on classical, 
relativistic field theories, but of course one could extend our results and conclusions to the Newtonian regime by taking the 
appropriate limit. For a brief discussion on the quantum aspects of these models see Section~\ref{discussion} at the end of this paper.

A perfect fluid is one with no dissipation effects~\cite{Landau}. For a perfect fluid in general relativity~\cite{Maartens:1996vi} 
the energy-momentum tensor and the entropy flux can be written in the form $T_{\mu\nu}=(\rho+p)u_{\mu}u_{\nu}+pg_{\mu\nu}$, and $S_\mu = s u_{\mu}$, respectively, with $\rho$, $p$ and $s$ the energy density, pressure, and entropy density measured by an observer at rest with respect to the fluid. The velocity $u^{\mu}$ is a four-vector pointing to the future, $u^0>0$, and normalized to $u_\mu u^\mu =-1$, with the spacetime metric $g_{\mu\nu}$ taking the mostly-plus-signature. Spacetime indexes are raised and lowered using the spacetime metric, e.g. $u_{\mu}=g_{\mu\nu}u^{\nu}$. Both quantities, the energy-momentum tensor and the flux of entropy are covariantly conserved for a perfect fluid, $\nabla_\mu T^{\mu\nu}=\nabla_\mu S^\mu=0$.

Additionally we can have other conserved currents, $\nabla_\mu N_i^\mu =0$, such as those associated to the particle or baryon numbers.
Here the letter $i$ is a label for these currents. For a perfect fluid the conserved currents are all parallel, and we can write $N_i^\mu = n_i u^\mu$, with $n_i$ a charge density. As usual, in order to close the system we need a relation of the form $\rho=\rho(s,n_i)$; ultimately it should be provided by the micro-physics. The other quantities of interest can be read from the thermodynamic relations, such as the temperature, $T=(\partial\rho/\partial s)_{n_i}$, the chemical potential, $\mu_i = (\partial\rho/\partial n_i)_{s,n_{j}}$, and so on. All these variables are related by the Euler equation, $\rho + p - Ts - \mu_i n_i =0$.

For the particular case of a potential flow, i.e. a fluid with no vorticity, $\Omega_{\mu\nu}=\nabla_\mu V_\nu - \nabla_\nu V_\mu =0$, 
we can always write $V_\mu = \partial_\mu\Phi$, with~$\Phi$ a velocity-potential and $V_{\mu} = v u_\mu$ the Taub current~\cite{Taub}. Here $v=h/s$ is a measure for the enthalpy per unit of entropy, and $h=\rho+p$ is the enthalpy density. For ``standard'' fluids (see for instance Ref.~\cite{Callen}) the Taub charge is positive-definite, $v>0$, with $v=0$ only in vacuum, i.e. $s=n_i=0$ for all $i$.

Leaving fluids aside, a local, minimally coupled to gravity, Lorentz invariant field theory with no more than two derivatives acting 
on a real field $\phi$ is described by the action 
\begin{equation}\label{action.scalar}
 S=\int d^4 x\sqrt{-g}\,M^4\mathcal{L}(\phi/M,X/M^4)\, .
\end{equation}
For our purposes $\phi$ is a Lorentz scalar measured in units of energy, and $X= -\frac{1}{2}\partial_\mu\phi\partial^{\mu}\phi$ is the kinetic 
term. We are taking units with $c=\hbar=1$, and $M$ is an energy scale. A scalar field described by an action of the form~(\ref{action.scalar}) is usually dubbed {\it k-essence}~\cite{k-essence}. In this language the Lagrangian density of a canonical scalar field takes the form $\mathcal{L}=X-M^4V(\phi/M)$, with $V$ a potential term. In order to include the dynamics of the gravitational interaction we only have to add a Einstein-Hilbert term to the expression above; however, for the purposes of this paper and with no loss of generality, we will restrict our attention to fixed background spacetime configurations.

Invariance under local Lorentz transformations defines energy and momentum. The energy-momentum tensor associated to the scalar field can be obtained varying the action in Eq.~(\ref{action.scalar}) with respect to the spacetime metric,
\begin{equation}\label{energymomentum.scalar}
 T_{\mu\nu} = \mathcal{L}' \partial_\mu\phi\partial_\nu\phi + \mathcal{L}g_{\mu\nu}\, .
\end{equation}
From now on and in order to simplify the notation we will omit the scale $M$. Here the prime denotes the derivative with respect to the kinetic term. Using the relations
\begin{equation}\label{identification.1}
 u_{\mu} = \frac{\partial_\mu \phi}{\sqrt{2X}}\, ,\quad 
 p = \mathcal{L} \, ,\quad 
 \rho = 2X\mathcal{L}'-\mathcal{L}\, ,
\end{equation}
we can identify the energy-momentum tensor of a scalar field with that of a perfect fluid, as long as the kinetic scalar is positive definite, $X>0$. In addition, if Eq.~(\ref{action.scalar}) is invariant under shift-transformations, $\phi\to\phi+\textrm{const.}$, that is, if the Lagrangian density does not depend explicitly on the scalar field, $\mathcal{L}=\mathcal{L}(X)$, we can also identify the Noether current 
\begin{equation}\label{noether}
 J_\mu = \mathcal{L}' \partial_\mu \phi
\end{equation}
with the flux of entropy in a perfect fluid, 
\begin{equation}\label{identification.2}
 s= \sqrt{2X} \mathcal{L}' \, .
\end{equation}
From the identities in Eqs.~(\ref{identification.1}) and~(\ref{identification.2}) we can read $v=\sqrt{2X}$. The energy-momentum tensor~(\ref{energymomentum.scalar}) and the Noether current~(\ref{noether}) are both covariantly conserved. Fields with no potential term are known as {\it purely-kinetics}, and have been considered for their possible role to the dark matter and/or dark energy problems~\cite{Scherrer}.

Associated to the Eq.~(\ref{action.scalar}) there are no other conserved charges apart from the energy, momentum, and Noether charge.
Then, if there were any other thermodynamic charges in the fluid, they should be distributed on a trivial configuration, i.e. the fluid should be isentropic, $\bar{s}_i= s/n_i =\textrm{const.}$; this guarantees a barotropic relation of the form $p=p(\rho)$. If on the contrary extra thermodynamic charges do not exist, e.g. a gas of photons, we can simply identify the Taub charge with the temperature in the fluid, $v=T$; see Euler equation. 

According to the arguments in the previous lines, as long as the kinetic term is positive definite, it seems possible to identify a shift-invariant scalar field $\phi$ with the velocity-potential of an isentropic, perfect, rotation-less fluid $\Phi$. But, what happens if the kinetic scalar changes sign? Naturally the identifications in Eqs.~(\ref{identification.1})~and~(\ref{identification.2}), and in particular that for the vector $u_\mu$, break down. If the dynamical evolution of the scalar field prevented a sign inversion, we could forget this concern. Something similar happens, for instance, when a perfect fluid is isentropic, or rotation-less, at a given instant of time, i.e. 
on a given Cauchy hypersurface: the dynamics maintains constant entropy per particle and no-vorticity along the fluid evolution. However, as we will find next by means of a simple example, this is not the case for the character of the derivative terms, Section~\ref{example}. In order to prevent this change of sign in the kinetic scalar a constraint should be introduced in the action. This is considered in Section~\ref{constraint}. We come back to our example in Section~\ref{revisiting}, and conclude in Section~\ref{discussion} with some comments and discussion.

\section{A simple example}\label{example}

Consider the case of a canonical scalar field with no potential term, $\mathcal{L}=X$. This theory is linear, and then it is easier to find exact solutions. It is also shift-invariant, and, according to the relations in Eqs.~(\ref{identification.1}) and~(\ref{identification.2}), it would seem possible to identify this field with the potential flow of a {\it stiff}, perfect fluid $p=\rho$.

In order to see that this identification is not always viable, let us look for the solutions of the form 
$\phi(t,x)=\varphi(t)+c_1x$ living on a flat, Robertson-Walker spacetime background, $ds^2 = -dt^2 + a^2(t)(dx^2+dy^2+dz^2)$, with $c_1$ constant and $a(t)$ the scale factor. (Remember that we are working in the test field approximation, that is, the scalar field does not gravitate and the function $a(t)$ is fixed {\it a priori}.)

From the Klein-Gordon equation, $\Box\phi=0$, we obtain
\begin{equation}\label{phi.t}
 \phi(t,x) = c_2 \int \frac{dt}{a^3 (t)} +c_1 x \, ,
\end{equation}
with $c_1$ and $c_2$ two arbitrary constants. We can use the family of two-parametric solutions in Eq.~(\ref{phi.t}) to construct 
the kinetic scalar as a function of the scale factor,
\begin{equation}
 X= \frac{1}{2} \left(\frac{c_2^2}{a^6}-\frac{c_1^2}{a^2} \right) \, . \label{solution}
\end{equation}
In order to identify the scalar field with the velocity-potential of a perfect fluid we need to satisfy $X>0$. From Eq.~(\ref{solution}) it is evident that this condition is verified at early times, when $a<a_{\textrm{crit}}=\sqrt{|c_2/c_1|}$ (the particular value of this quantity depends on the initial conditions for the scalar field). However, for $a \geq a_{\textrm{crit}}$, the inequality $X>0$ does no longer holds. 

The moral is simple: not all the particular solutions to a shift-invariant scalar field satisfy the condition necessary to mimic a perfect fluid, $X>0$. But even if they do at a certain initial time, $t_0$, the evolution of the system can change this behavior. Imagine for instance a universe filled with a scalar field, like in the inflationary model, and assume that this field is invariant under shift-transformations. If there were some small perturbations to the homogeneous and isotropic background, we could not guarantee a perfect-fluid-solution, even if the identification of the scalar field with a perfect fluid were possible in the early universe. But, how is it possible that something that looks like a perfect fluid, and evolves like a perfect fluid, reaches a state that does {\it not} look like a perfect fluid?

\section{The constraints}\label{constraint}

In order to improve our understanding of the previous section, we find it appropriate to start from the action principle describing a 
perfect fluid in general relativity. Following Schutz formalism \cite{Schutz} (see also Refs.~\cite{resto}, and Ref.~\cite{Newtonian} for a Newtonian description), the Lagrangian density of an isentropic, rotation-less perfect fluid with equation of state $\rho=\rho(s, n_i)$ is given by 
\begin{equation}\label{action.fluid}
 S=\int d^4x \sqrt{-g}\left[ -\rho(s, n_i) - S^\mu\partial_\mu\Phi +\lambda(S_\mu S^\mu+s^2)\right]\, .
\end{equation}
Eq.~(\ref{action.fluid}) is a functional of the spacetime metric, $g_{\mu\nu}$, the entropy flux, $S^\mu$, the velocity-potential, $\Phi$, the entropy density, $s$, and a new field $\lambda$. For an isentropic fluid, $\bar{s}_i=\textrm{const.}$, the charge density
$n_i$ is not a variable anymore, i.e. $n_i=\textrm{const.}\times s$. Here $\lambda$ is a Lagrange multiplier that guarantees the standard normalization for the flux of entropy, $S_\mu S^\mu=-s^2$. (The necessity of constraints is analyzed in full detail in a seminal paper by Schutz and Sorkin, Ref.~\cite{SchutzSorkin}.) At this point we can also look at the velocity-potential $\Phi$ as a Lagrange multiplier, necessary to guarantee entropy conservation: integrating by parts and removing a surface term we can replace $S^\mu\partial_\mu\Phi$ by $\Phi\nabla_{\mu} S^\mu$ in Eq.~(\ref{action.fluid}). (For a more general, {\it rotational} fluid, the Lagrangian density requires an extra term of the form $S^\mu\beta_A\partial_\mu \alpha^A$, with $\alpha^A$ and $\beta_A$ additional fields related to the Lagrangian coordinates of the fluid, $A=1,2,3$; see the previous references for further details.)

Varying Eq.~(\ref{action.fluid}) with respect to the spacetime metric, and using the thermodynamic relations, we obtain the 
energy-momentum tensor of a perfect fluid. The variation with respect to the Lagrange multipliers $\Phi$ and $\lambda$ give the equations for the conservation, $\nabla_\mu S^\mu = 0$, and normalization, $S_{\mu} S^\mu =- s^2$, of the flux of entropy, respectively.
Finally, varying with respect to $S_{\mu}$ and $s$ we get
\begin{subequations}
\begin{eqnarray}
 u_{\mu} &=& (1/v) \partial_\mu \Phi \, , \label{u.pot}\\
 \lambda &=& v/2s \, . \label{lambda}
\end{eqnarray}
\end{subequations}
Eq.~(\ref{u.pot}) is a decomposition (known as the Clebsch representation) of $u_\mu$ in terms of the velocity-potential; compare with the first identity in Eq.~(\ref{identification.1}). Since $V_{\mu} = \partial_\mu\Phi$, this guarantees a fluid with no vorticity, as we anticipated at the beginning of this section. Finally, Eq.~(\ref{lambda}) gives the evolution of the field $\lambda$ in terms of the Taub charge and the entropy density in the fluid.

When the equations of motion hold, Eq.~(\ref{action.fluid}) reduces to
\begin{subequations}
\begin{equation}\label{action.pt}
 S_{\textrm{on-shell}}=\int d^4x \sqrt{-g}\, p(v,\bar{s}_i=\textrm{const.})\, ,
\end{equation}
with the square of the Taub charge 
\begin{equation}\label{eq.2}
v^2=-\partial_{\mu}\Phi\partial^{\mu}\Phi
\end{equation}
\end{subequations}
a kind of kinetic scalar. We can read this last identity from Eq.~(\ref{u.pot}), using $u_\mu u^\mu =-1$. The Lagrangian density {\it on-shell} coincides with the pressure in the fluid, Eq.~(\ref{action.pt}), and we can arrive to the energy density from the thermodynamic relations, $\rho=(\partial p /\partial v)_{\bar{s}_i}v-p$; compare with the second and last identities in Eq.~(\ref{identification.1}). The definition of the Taub charge, $v=h/s$, reproduces the identity in Eq.~(\ref{identification.2}).

Since $v^2$ is positive definite, the derivative terms cannot change character, i.e. if they were born time-like, they will remain 
time-like along fluid evolution. Eq.~(\ref{action.pt}) is the action for a shift-invariant, k-essence-like velocity-potential, with the kinetic scalar measuring the Taub charge in the fluid. We are then lead to the same identifications as those reported in the Introduction, but now starting from the action principle describing a perfect fluid in general relativity.

\section{Revisiting the example}\label{revisiting}

\begin{figure}[t]
\centering
\includegraphics[width=0.45\textwidth]{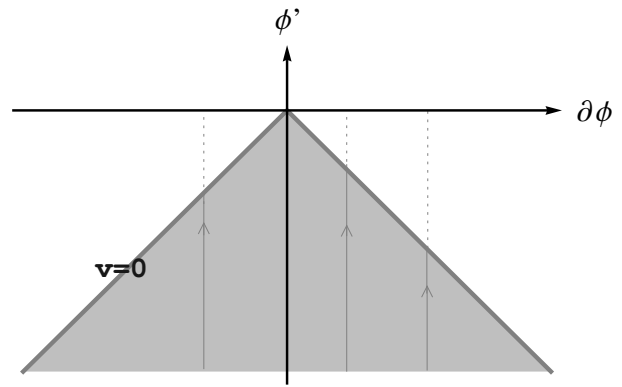}
\caption{
Shady region: values of $\phi ' = \partial\phi(\eta,\vec{x})/\partial\eta$ and $\partial\phi = \partial\phi(\eta,\vec{x})/\partial x$
(at fixed $\vec{x}$) that allow a perfect fluid description, with $d\eta=dt/a$ a conformal time.
Solid lines with an arrow represent perfect fluid evolution. At $v=0$ the evolution of the scalar field
and the perfect fluid branch off: dashed lines represent scalar field evolution with no perfect fluid
analogue.
}
\label{configuration}
\end{figure}

Let us come back to the example in Section~\ref{example}. There, at time $t_0$ we can fix $\phi(t_0, \vec{x})$ and $\dot{\phi}(t_0, \vec{x})$ arbitrarily. This is no longer true for the velocity-potential of a perfect fluid, where we can choose $\Phi(t_0, \vec{x})$ and $v(t_0,\vec{x})>0$, i.e. the 3-velocity and the Taub charge on a Cauchy hypersurface, but the constraint in Eq.~(\ref{eq.2}) fixes 
\begin{equation}
 \dot{\Phi}(t_0, \vec{x}) = - \left[ v^2(t_0, \vec{x}) + \frac{1}{a^2(t_0)}\partial_i^2\Phi(t_0, \vec{x})\right]^{1/2} \, .
\end{equation}
The minus sign in the square root is chosen to guarantee $u^0 >0$. The points in phase space that admit a perfect fluid description are restricted, see Fig.~\ref{configuration}. But even if we start from a state in the subspace that allows such a description, shady region in Fig.~\ref{configuration}, the dynamics seems to bring the system into the space that does not admit a perfect fluid analogue. Is then the dynamics of the scalar field different to that of a perfect fluid?

Varying Eq.~(\ref{action.pt}) with respect to the velocity-potential, we obtain
\begin{equation}\label{eq.1}
 \nabla_\mu \left[\frac{1}{v}\left(\frac{\partial p}{\partial v}\right) \partial^\mu \Phi \right] =0 \, .
\end{equation}
In order to identify Eq.~(\ref{eq.1}) with a shift-invariant, Klein-Gordon equation, we must satisfy (we are using the notation in the 
Introduction)
\begin{equation}\label{equality}
 \frac{1}{v}\left(\frac{\partial p}{\partial v}\right) = \frac{\partial\mathcal{L}}{\partial X} \, .
\end{equation}
This identification is possible as long as $v > 0$. For standard fluids a vanishing Taub charge, $v=0$, means vacuum (here we are talking about a classical vacuum). There is nothing beyond the vacuum of a perfect fluid, and the state of the system freeze down at that point in the evolution. In the context of the example in Sec.~\ref{example}, the expansion dilutes the matter content in the universe; additionally, if the velocity of the matter fields with respect to the expansion does not vanish, $c_1\neq 0$, it is possible to leak out the fluid at finite cosmological time. 

As it is natural from Fig.~\ref{configuration}, in order to reach the region in phase space that does not admit a perfect fluid description, the scalar field should go through the ``vacuum divide'', $v=0$. At that point the Klein-Gordon equation does not describe the dynamics of a perfect fluid anymore, and the evolution of the two systems, the perfect fluid and the scalar field, branches off: whereas the perfect fluid remains in vacuum, the scalar field follows an evolution that does not admit a perfect fluid analogue, seeping through a region with ``imaginary Taub charge''.

\section{Discussion}\label{discussion}

The identification of a scalar field with the velocity-potential of a perfect fluid is possible, as long as the scalar field is shift-invariant, $\phi\to\phi+\textrm{const.}$, and the perfect fluid isentropic and rotation-less. However, actually not all the scalar field solutions mimic hydrodynamic motion: only those that satisfy the constraint $X>0$ verify this identification. In general, for solutions with some space dependency, the scalar field only mimics a perfect fluid for a finite period in the evolution. After that period of time, the evolution of the two systems, the scalar field and the perfect fluid, branches off. In terms of the example we considered in Section~\ref{example}, the perfect fluid reaches the vacuum state at finite cosmological time (unless $c_1=0$), and it freezes down at that point. From there on, the scalar field develops anisotropic configurations with no perfect fluid analogue, and the two systems start to differ, see Fig.~\ref{configuration}. (For a discussion on anisotropic scalar field configurations please see Ref.~\cite{Diez-Tejedor:2013sza}.)

If the scalar field is not invariant under shift-transformations, the identification with a perfect fluid is no longer possible. Even though the relations in Eq.~(\ref{identification.1}) are still allowed, there are no other conserved charges apart form the energy and momentum to identify with the entropy density. Then, in general, the dynamics of a scalar field and a perfect fluid differ. 

However, a {\it formal} relation can be sometimes useful. For instance, for static configurations the character of the derivative terms does not change. Then, if they are time-like, we can use Eq.~(\ref{identification.1}) to identify a pressure and an energy density, and in some cases with high symmetry even to obtain an effective barotropic relation of the form $p=p(\rho)$. Something similar happens for the cosmological homogeneous and isotropic solutions, where there is time evolution, but no spatial gradients. However, it is important to remark that the existence of a relation of the form $p(\rho)$ for particular solutions does {\it not} imply an ``equation of state'' for the scalar field, and, contrary to what happens in these two cases above (and others considered in the literature), the identifications reported in this note are general, and not background-dependent.

Let us conclude with some words about the quantum regime of the scalar/perfect fluid field models considered in this paper. The quantization of a canonical scalar field is well understood, in flat~\cite{flat} as well as in curved~\cite{curved} spacetimes; however, noncanonical scalar fields are definitely more subtle. On the hydrodynamic side, one could probably argue that the noncanonical fields describing the collective modes of a perfect fluid are restricted to the classical world, and only the small perturbations (``phonons'') around their background values are subjected to quantization; see for instance Refs.~\cite{Hu} for a discussion along these lines in the context of quantum gravity. A quantum description of a perfect fluid at zero temperature has been recently considered in Ref.~\cite{Nicolis}; see also Refs.~\cite{others}. Using an effective field theory approach, the authors in Ref.~\cite{Nicolis} identify a number of interesting aspects (e.g. strong coupling at low energies, an analog of Coleman's theorem) that could be relevant for the description of perfect fluids at very low temperatures, when the thermal effects are still sub-dominant. Note however that again the quantization of the perfect fluid is carried out in the canonical (perturbative) way. This discussion is however beyond the scope of this paper, and we refer the interested reader to \cite{Hu,Nicolis,others} for further details.

\begin{acknowledgments}
   
We are grateful to Alex Feinstein, Alma X. Gonzalez-Morales, Dar\'io N\'u\~nez and Olivier Sarbach for useful comments and discussions.
This work was partially supported by PIFI, PROMEP, DAIP-UG, CAIP-UG, the ``Instituto Avanzado de Cosmolog\'ia'' (IAC) collaboration, 
and CONACyT M\'exico under grants 182445 and 167335.

\end{acknowledgments}

\end{document}